\documentstyle[amsfonts,eqsecnum,prd,aps,preprint]{revtex}
\tightenlines

\def\buildchar#1#2#3{\null \! \mathop {\vphantom {#1}\smash
#1}\limits ^{#2}_{#3}\!\null }
\def\OT#1{\buildchar{{#1}}{\;_\sim}{}\/}
\def\UT#1{\buildchar{{#1}}{}{^\sim}\/}
\def\OTT#1{\buildchar{{#1}}{\;_\approx}{}\/}

\begin{document}
\draft

%234567890 234567890 234567890 234567890 234567890 234567890 234567890

\title{Quantum general invariance and loop gravity}

\author{D.\ C.\ Salisbury
\footnote[2]{IARD 2000 Conference Proceedings,
Tel Aviv, 2000}}
\address{Department of Physics,
Austin College, Sherman, Texas 75090-4440, USA, dsalisbury@austinc.edu}

%\date{24 April 2001}
\maketitle

\begin{abstract}
A quantum physical projector is proposed for generally covariant
theories which are derivable from a Lagrangian. The projector is
the quantum analogue of the integral over the generators of finite
one-parameter subgroups of the
gauge symmetry
transformations which are connected to the identity. Gauge
variables are retained in this formalism, thus permitting the
construction of spacetime area and volume operators in a
tentative spacetime loop formulation of quantum general
relativity.
\end{abstract}

\pacs{04.20.Fy, 11.10.Ef, 11.15.-q \hfill gr-qc/0105097}

\section{Introduction}

In a recent series of papers I and my collaborators Josep Pons and
Larry Shepley have analyzed the classical gauge symmetries in the
phase space (cotangent bundle) formulation of a wide class of
generally covariant dynamical systems which are derivable from a
Lagrangian
\cite{pons/salisbury/shepley/97,pons/salisbury/shepley/99a,pons/salisbury/shepley/99b,pons/salisbury/shepley/99c}.

Our point of departure is the requirement that symmetries present
in the configuration-velocity (tangent bundle) approach must be
projectable onto the cotangent bundle. This view was suggested to
us by pioneering work of Lee and Wald \cite{lee/wald/90} which
with some modifications provides a theoretical framework for
symmetry explorations which were initiated by Bergmann and Komar
\cite{bergmann/komar/72}, and extended by myself and Sundermeyer
\cite{salisbury/sundermeyer/83a,salisbury/sundermeyer/83b}. The
outcome of our analysis is that in general a projectable gauge
symmetry group exists, and it is a transformation group on the
configuration-velocity variables. The group transformations arise
from spacetime diffeomorphisms which contain a compulsory
dependence on the lapse and shift gauge functions. For those
dynamical models which possess additional symmetries, such as
Ashtekar's formulation of general relativity, one of these
additional gauge transformations must be added to the
diffeomorphism-induced transformations to achieve projectability.
We have constructed the complete symmetry generators in the
constrained phase space which contains all of the dynamical
variables, including the gauge variables.

Two features of the resulting phase space formulation are
especially significant, and both were noted almost thirty years
ago by Bergmann and Komar \cite{bergmann/komar/72}: First, rigid
time translation, i.e., evolution in time, is {\it not} a gauge
symmetry. Second, because of the fact that the gauge algebra
contains derivatives of the metric, true symmetry invariants will
be non-local. This occurs since nested commutators of generators
contain derivatives of arbitrarily high order.

The decoupling of symmetry and time evolution has profound
implications. In this essay I shall further elaborate on a recent
proposal which exploits the true symmetry in constructing quantum
invariants in loop gravity \cite{pons/salisbury/shepley/00}. The
non-local nature of the invariants will be manifest. The basic
idea is to retain gauge variables as quantum variables, but in a
novel fashion: Since the classical gauge variables are arbitrary
functions of time, we shall interpret both spatial {\it and time}
coordinates of gauge variables as indices. In other words, gauge
field variables each constitute a $4\times \infty$ set.

In Section II I will give a physical motivation for retaining the
gauge variables. Section III is an overview of the classical
symmetry structure, culminating in a presentation of the finite
classical generator of one-parameter subgroups of symmetry transformations.
This object is
employed in Section IV in the construction of a physical quantum
projector.  Finally, in
Section V I turn to the loop formulation of quantum gravity. I
propose a larger kinematical Hilbert Space formed not only from
spatial Wilson loops with their associated holonomies, but also
SU(2) gauge invariant loops containing legs in the new parameter
time directions. The resulting structure permits the construction
of true quantum spacetime invariants.

\section{Physical Motivation}

One might be tempted to think that a foliation of spacetime into
fixed time slices would irretrievably destroy four-dimensional
spacetime symmetry. Indeed, most approaches to quantum gravity are
content with exploring the consequences of the residual spatial
diffeomorphism and internal gauge group symmetries. The so-called
scalar or Hamiltonian constraint is recognized as a generator of
time evolution, although time evolution is generalized to
incorporate the notion of advance of a ``multifingered time'', an
idea that has been promoted by Kucha\v{r} and others. (See
\cite{isham/92} for a review of the problem of time in general
relativity.) Multifingered time advances along directions
perpendicular to the constant time hypersurfaces. Our work
supplies an explanation - first noted by Lee and Wald
\cite{lee/wald/90} - for this strange dependence of
diffeomorphisms on the spacetime metric; it is required to achieve
projectability under the Legendre map to phase space. The earlier
work by myself and Sundermeyer provides an equivalent explanation,
as explained in \cite{pons/salisbury/shepley/97}. Multifingered
advance in time is in fact a gauge {\it group} symmetry, and it
is this recognition that a spacetime diffeomorphism-induced {\it
group} symmetry remains in the phase space formulation of general
relativity that has neither been recognized, nor exploited in
quantum general relativity.

What might we hope to gain from this larger symmetry? First of
all, since rigid time evolution is not a symmetry we can
reasonably expect that time will not be ``frozen''. Different
times are in principle distinguishable, even in the context of
vacuum gravity. Second, although the classical gauge functions
are freely prescribable functions of time, they do undergo quite
specific, known symmetry transformations. Spacetime invariants,
obtained by applying these transformations to functionals
containing them will depend on them in a highly nontrivial
manner. This has important implications for quantum gravity both
in the conventional Wheeler-DeWitt treatment, and for newer loop
versions employing the Ashtekar connection. The gauge variables
must be retained as quantum operators.

In fact, we ought to {\it insist} that the lapse and shift be
retained; a quantum ``fuzziness'' only in spatial proper distances
manifestly breaks the underlying symmetry of the theory. In
retaining the gauge variables it becomes possible to construct
operators representing true spacetime distances, areas and
volumes. The absence of such objects in the loop approach is
especially problematical, given the emergence of discreteness in
spatial areas and volumes \cite{rovelli/smolin/95a}.

\section{Classical Gauge Symmetry}

My collaborators and I have investigated conditions that must be
fulfilled by infinitesmal gauge symmetry transformations in the
original classical Lagrangian formalism of a wide variety of
generally covariant theories in order that these variations can be
mapped under the Legendre map to phase space. The theories include
the relativistic particle, the relativistic string, conventional
general relativity \cite{pons/salisbury/shepley/97} ,
Einstein-Yang-Mills \cite{pons/salisbury/shepley/99a}, a real
triad version \cite{pons/salisbury/shepley/99b}, and the Ashtekar
formulation, of general relativity
\cite{pons/salisbury/shepley/99c}. The resulting projectable
infinitesmal symmetry generators $G[\xi;t]$ all have the
following structure:
\begin{equation}
    G[\xi;t] = P_{A} \dot\xi^{A}
       + ({\cal H}_{A}
       + P_{C''}N^{B'}{\cal C}^{C''}_{AB'})\xi^{A}\ ,
           \label{GGGGG}
\end{equation}
where the structure functions are obtained from the closed Poisson
bracket algebra
\begin{equation}
    \{ {\cal H}_{A},{\cal H}_{B'} \}
    =: {\cal C}^{C''}_{AB'} {\cal H}_{C''}\ ,
\end{equation}
and where spatial integrations at time $t$ over corresponding
repeated capital indices are assumed. The $ N^A$ are the gauge
functions. Their canonical momenta $P_A$ are primary constraints.
The physical phase space is further constrained by secondary
constraints ${\cal H}_A$. These constraints generate symmetry
variations of the non-gauge variables. The ``descriptors'' $\xi^A$
are arbitrary spacetime functions.

If there is no symmetry in the Lagrangian description beyond
general covariance, the indices $A$ range from zero to three. The
corresponding gauge functions are the lapse $N$ and the shift
$N^a$, so $ N^A = \{N,N^a\}$. (My index convention is that spatial
indices are lower-case latin letters from the beginning of the
alphabet.) The lapse and shift appear in the spacetime metric
\begin{equation}
    (g_{\mu\nu}) = \left(
        \begin{array}{cc}
            -N^{2}+N^{c}N^{d}g_{cd} & g_{ac}N^{c}  \\
            g_{bd}N^{d} & g_{ab}
        \end{array}
                \right) \ .
    \label{eq:gmunu}
\end{equation}
The projectable infinitesmal symmetries are induced by spacetime
infinitesmal diffeomorphisms of the form
\begin{equation}
    x'^\mu = x^\mu -\delta^\mu_a \xi^a - n^\mu \xi^0.
    \label{diff}
\end{equation}
The normal $ n^\mu$ to the fixed time hypersurface is expressed as
follows in terms of the lapse and shift:
\begin{equation}
    n^\mu = (N^{-1},-N^{-1}N^a).
\end{equation}

Projectable configuration-velocity functions may not depend on
time derivatives of the lapse and shift; equation (\ref{diff})
represents the most general infinitesmal diffeomorphism producing
variations which satisfy this requirement.

If gauge symmetries exist beyond those induced by diffeomorphisms
one obtains additional projectability conditions. I will call
these additional symmetries ``internal symmetries''. In all such
theories we have considered, a Yang-Mills type connection
constitutes an additional configuration variable. The Lagrangian
does not depend on the time derivative of the temporal component
of this connection, hence the additional projectability
requirement is that symmetry variations may not depend on time
derivatives of this temporal component. In this case the index $A$
acquires an additional range, over the dimension of the Lie
algebra of the additional gauge group.

Specifically, in Ashtekar's formulation of general relativity the
Ashtekar connection $A^i_\mu$ is an element of the Lie algebra
$so(3,R)$ or $so(3,C)$, and the lower case latin indices $i$
range from one to three. It turns out that variations of the
temporal component of the connection induced by the
diffeomorphisms (\ref{diff}) depend on time derivatives of this
component, and hence these variations are {\it not}
projectable. Projectable infinitesmal variations are obtained by
adding an internal gauge transformation constructed from the
diffeomorphism descriptor $\xi^0$ and the connection form
contracted with the hypersurface normal. In the complex Ashtekar
case the required internal gauge descriptor is $A^i_\mu n^\mu
\xi^0-i N^{-1} T^{ai} N_{,a} \xi^0$. ($T^{ai}$ are the triad
fields, from which we obtain the contravariant spatial metric
$e^{ab} = T^{ai}T^{bi}$). In all of the cases we have considered,
the constraint ${\cal H}_0$ generates the corresponding
projectable infinitesmal variation of the non-gauge variables.

Returning to the infinitesmal generator (\ref{GGGGG}), let me
complete the list of variables in the Ashtekar case. The gauge
functions are $\{\UT N, N^a , -A^i_0\}=: N^A$, with their
canonical momenta, which are primary constraints: $\{\OTT P,\OT
P_a,-\OT P_i\}=: P_A$. $\OT P_i$ is the momentum conjugate to
$A^i_0$. The secondary constraints are $\{\OTT{\cal  H}_0,\OT{\cal
H}_a,\OT{\cal H}_i\}=:{\cal H}_A$, where $\OT{\cal H}_i$ generates
internal $SO(3)$ rotations of the non-gauge variables. (As has now
become conventional, densities of arbitrary positive weight under
spatial diffeomorphisms are represented by an appropriate number
of tildes over the symbol. For negative weights the tilde is
placed below the symbol.) The momenta conjugate to $A^i_a$ are
the densitized triad field $\OT T^a_i := det(t^i_a) T^a_i$ where
$t^i_a$ is the covariant triad, the inverse of $T^a_i$: $t^i_a
T^a_j = \delta^i_j$. (In the complex case the canonical pair is
actually $\{\OT T^{a}_{i}, i A^i_a\}$.)

It must be stressed that the generator $G[\xi;t]$ in (\ref{GGGGG})
is actually a function of the time $t$, and it is assumed that the
canonical variables appearing in this expression are solutions of
the equations of motion. The gauge functions are however almost
arbitrary; the only condition on them is that the lapse must be
strictly positive. Also, the primary constraints undergo a
trivial evolution; they are always zero. The evolution of the
non-gauge variables, on the other hand, is generated by the
canonical Hamiltonian, where explicit choices are made for the
gauge functions.

The canonical Hamiltonian is $H_c = N^A{\cal H}_{A}$. It generates
time evolution of the non-gauge variables. We do not alter either
the equations of motion or gauge variations in recognizing that
since the gauge variables $N^A$ are arbitrary functions of both
space and time we can add a term $\int^{\infty}_{-\infty} d^{4}x
P_{A}(x) \dot N^{A}(x)$ to the canonical Hamiltonian. The new
Hamiltonian becomes
\begin{equation}
        H(t) = \int^{\infty}_{-\infty} dt'\int d^{3}x P_{A}(\vec x,t')
    \dot N^{A}(\vec x,t') + \int d^{3}x N^A(\vec x,t){\cal H}_{A}.
    \label{Ht}
\end{equation}
Thus the classical evolution of all dynamical variables is
effected by the time-ordered evolution operator
\begin{equation}
    \{ - ,U[t,t_0] \} = {\cal T} \exp\left(\int^{t}_{t_{0}}
dt' \, \{ - , H(t')\}\right)\ ,
    \label{canevol}
\end{equation}
where $\{ , \}$ represents the Poisson Bracket which I here
generalize to include an integral over the time indices of the
gauge variables;
\begin{equation}
    \{ N_{A}(\vec x, t), P_{B}(\vec x', t')     \} = \delta^{3}(\vec
x,\vec
    x') \delta (t,t') .
\end{equation}

Let me demonstrate that $U[t,t_0]$ does indeed correctly rigidly
translate the gauge functions $N^A$ in time. We have
\begin{eqnarray}
        \{N^{A}(t_{0}), U[t,t_0] \} &=& N^{A}(t_{0}) + \int^{t}_{t_{0}}
        dt_{1}\{N^{A}(t_{0}),
        \int dt' P_{B}(t') \dot N^{B}(t' \} \nonumber \\
        && + \int^{t}_{t_{0}} dt_{2}
        \int^{t_{2}}_{t_{0}} dt_{1} \{ \{N^{A}(t_{0}), \int dt''
        P_{B}(t'') \dot
        N^{B}(t'') \},\int dt' P_{C}(t') \dot N^{C}(t') \} + \ldots
        \nonumber \\
        &=& N^{A}(t_{0}) + \dot N^{A}(t_{0}) (t - t_{0}) +
        {1 \over 2} \ddot N^{A}(t_{0}) (t - t_{0})^{2} + \ldots
        \nonumber \\
        &=&\sum_{n=0}^{\infty}{1 \over n!}{d^{n} N^{A}(t_{0}) \over
        d t^{n}} (t-t_{0})^{n} = N^{A}(t).
\end{eqnarray}
(In the first equality we used the fact that the time-dependent
contribution to the second term in the generator (\ref{Ht}) yields a
vanishing Poisson bracket. Therefore the integration over the times
$t_{i}$ is trivial.)
Also $U(t,t_{0})$ maintains the primary constraint $P_{A} \approx 0$.

Our next task is to find the expanded infinitesmal generator,
expressed in terms of the larger set of canonical gauge
variables, which effects the correct variations of the canonical
variables. The following object does the job:
\begin{equation}
     G_{e}[\xi;t] = \int dt' P_{A}(t') \dot\xi^{A}(t')
       + {\cal H}_{A}(t)\xi^{A}(t)
       +\int dt' P_{C''}(t')(N^{B'}(t'){\cal
       C}^{C''}_{AB'}(t')\xi^{A}(t')\ .
\end{equation}

We turn now to the construction of finite one-parameter subgroups
of the full gauge transformation group. We will build them up as
usual from infinitesmal gauge transformations. We begin with the
more familiar three-dimensional spatial diffeomorphism group. For
this purpose it is instructive to display explicitly the general
infinitesmal symmetry variations of the lapse and shift gauge
variables:
\begin{mathletters}\label{d}
\begin{eqnarray}
    \delta N & = & \dot\xi^{0} + \xi^{a}N_{,a} -N^{a}\xi^{0}_{,a} \ ,
                                                      \label{d.N}  \\
    \delta N^{a} & = & \dot\xi^{a}
        - Ne^{ab}\xi^{0}_{,b} + N_{,b}e^{ab}\xi^{0}
                                                \nonumber \\
        && \qquad + N^{a}_{,b}\xi^{b}
        -N^{b}\xi^{a}_{,b} \  .
                                                     \label{d.Na}
\end{eqnarray}
\end{mathletters}%
If the descriptor $\xi^{0}$ vanishes the $\xi^{a}$ represent, for
a fixed time, vector fields on the three-dimensional spatial
manifold. They may be interpreted as tangents to a one parameter
family of spatial manifold maps. We build up the finite maps by
solving the set of ordinary differential equations which follow
from (\ref{diff}), taking the spatial descriptors to be $ds
\xi^a$ and setting $\xi^0 = 0$,
\begin{equation}
{dx^a \over ds} = -\xi^a (x). \label{xs}
\end{equation}
Using (\ref{xs}) it is straightforward to build up a formal power
series in $s$, since $\ddot x^a =\xi^a_{,b}\xi^b $, etc. The
result is
\begin{equation}
x^a(s) = \sum_n {(-1)^n s^n \over n !}\left(\prod_{i=1}^{n-1}
\xi^{a_i}\partial_{a_i}\right) \xi^a.
\end{equation}

The corresponding  formal generator of finite variations of the
dynamical variables is constructed with the aid of the
infinitesmal generator \cite{pons/salisbury/shepley/99c}
\begin{eqnarray}
G:&=&G_{e}[\xi^a,\xi^0=0,\xi^i=0;t] = \int dt' \OTT P_{a}(t')
\dot\xi^{a}(t')
       + {\cal H}_{a}(t) \xi^{a}(t)
       +\int dt' P_{C''}(t') N^{B'}(t')
       {\cal C}^{C''}_{aB'}(t')\xi^{a}(t') \nonumber \\
       &=&  \OT{\cal H}_{a}(t)\xi^{a}(t)
          +\int dt'\OTT P(t') \left(\UT N_{,a}(t')\xi^a(t') - \UT N(t')
\xi^a_{,a}(t')\right)
              \nonumber \\
    &&\qquad    +\int dt' \OT P_a(t') \left( N^a_{,b}(t') \xi^b(t')
                   - N^b(t') \xi^a_{,b}(t') \right)
                       \nonumber \\
    &&\qquad    + \int dt'\OT P_i(t')\left(F^i_{ab}(t')\xi^a(t) N^b(t')
          +i F^{ij}_{ab}(t') \OT T^{b}_{j}(t') \UT N(t')
          \xi^{a}(t')\right).
\end{eqnarray}
The finite generator is
\begin{equation}
1 + s \{- ,G\} + {s^2 \over 2!} \{ \{- ,G\},G\} + \ldots =
exp(s\{-,G\}).
\end{equation}

We now consider the situation when  the descriptors $\xi^0$ do
not vanish. The analysis of the finite one-parameter subgroups of
the full four-dimensional diffeomorphism-related gauge symmetry
group is substantially complicated by the fact that the group
depends on the dynamical variables. To simplify the discussion I
will consider as an example a non-field theoretical model, the
relativistic free point particle. Most convenient for our
purposes is a classical formulation with an auxiliary gauge
variable \cite{pons/salisbury/shepley/97}.

The relativistic free particle with mass one is described by the
Lagrangian
\begin{equation}  L = \frac{1}{2N} \dot x^\mu \dot x^\nu \eta_{\mu\nu}
    - \frac{1}{2} N,
        \label{partlag}
\end{equation}
where ${ x^{\mu}(\theta)}$ is the vector variable in Minkowski
spacetime, with metric $(\eta_{\mu\nu})={\rm{}diag}(-1,1,1,1)$,
and $N$ is an auxiliary variable whose equation of motion gives $N
= (-{\dot x}^\mu {\dot x}_\mu)^{1/2}$.  $N$ may be interpreted as
a lapse, with corresponding metric $g_{0 0}=-N^2$ on the manifold parametrized by
$\theta$.

This is a generally covariant model as the dynamics does not
change its form under arbitrary reparametrizations $\theta' =
\theta'(\theta)$. There exists a primary constraint $\pi \simeq
0$, and a secondary constraint $H = \frac{1}{2} ( p^\mu p_\mu +
1) \simeq 0$. The canonical Hamiltonian is $ H_c = \frac{N}{2}
( p^\mu p_\mu + 1)$. The projectable infinitesmal
reparametrizations are $\theta' =\theta - N^{-1} \xi^0$, and
according to (\ref{GGGGG}) the corresponding generator is
\begin{equation}
    G[\xi^{0};\theta] = \int d\theta'{\dot \xi^{0}(\theta')} \pi(\theta') +
    \xi^{0} (\theta) \frac{1}{2} (p_\mu(\theta) p^\mu(\theta) +1).
\end{equation}

Now referring again to (\ref{diff}) applied to the particle model,
we deduce that
\begin{equation}
{d\theta \over ds}|_{s=0} = -\xi^0 (\theta)N^{-1}(\theta)|_{s=0}.
\label{dthetads}
\end{equation}
Unfortunately, both $\xi^0$ and $N$ will alter their functional
dependence on $\theta$ under the one-parameter group, so
(\ref{dthetads}) is not particularly useful. We need to determine
directly the one-parameter family of transformations of lapses
$N_s(\theta)$ which is generated by the lapse-dependent
reparametrizations! First we note from (\ref{d.N}) that
\begin{equation}
{\partial N_s \over \partial s}(\theta) |_{s=0} = {\partial
\xi_s^0 \over \partial \theta}(\theta) |_{s=0}.
\end{equation}

We must be careful in writing  down the appropriate differential
equation for arbitrary s; the function $\xi^0$ also undergoes a
variation under this reparametrization. This occurs because there
is an essential difference between the metric-independent spatial
coordinate transformation (\ref{diff}) when $\xi^0$ vanishes, and
the spacetime transformation resulting from a nonvanishing
$\xi^0$. In the former case the descriptors $\xi^a$ are
invariant; the Lie derivative of $\xi^a$ with respect to itself
is zero. In the later case it is $n^\mu \xi^0$ which is
invariant. It follows from the variation of $N$ that $\xi^0$
transforms as a scalar. In fact, $\xi^0$ acquires a dependence on
$N$. We notice that the infinitesmal variation of $\xi^0$ under
the infinitesmal reparametrization $\theta' = \theta - ds \xi^0
N^{-1}$ is $\delta \xi^{0} = {\partial \xi^{0} \over d\theta } ds
\xi^0 N^{-1}$, so
\begin{equation}
{\partial \xi_s^0 \over \partial s} |_{s=0} = {\partial \xi_s^{0}
\over \partial\theta }  \xi^0 N^{-1} |_{s=0}.
\end{equation}
The one-parameter subgroup differential relations may now be
generalized to arbitrary parameter value $s$:
\begin{equation}
    {\partial N_s \over \partial s}(\theta)  = {\partial \xi_s^0
\over \partial \theta}(\theta),  \label{dNdss}
\end{equation}
and
\begin{equation}
{\partial \xi_s^0 \over \partial s}(\theta)  = {\partial \xi_s^{0}
\over \partial\theta }(\theta) \xi_s^0(\theta) N_{s}(\theta)^{-1} .
\label{dxidss}
\end{equation}

We now develop a formal power series solution in the parameter $s$
for $\xi_s^0 (\theta)$ and $N_s(\theta)$. We deduce from
(\ref{dNdss}) and (\ref{dxidss}) that ${\partial \over
\partial s}(N_s^{-1} \xi_s^{0}) = 0$. (This is simply the invariance of
$n^\mu \xi^0$ in this model.) Repeated use of this identity
results in the following expression for the n'th derivative of
$\xi^{0}$ with respect to $s$:
\begin{equation}
\xi_s^{0\,[n]} := {\partial^n \xi_s^0 \over \partial s^n}|_{s=0}=
\xi_{0} N^{-1} {d \over d\theta }\left( \xi^{0} N^{-1} {d \over
d\theta }\left( \xi^{0} N^{-1}{d \over d\theta } \left(\dots {d
\over d\theta } \left(\xi^{0} N^{-1}{d \xi^{0} \over d\theta
}\right) \ldots\right) \right)\right), \label{xi0n}
\end{equation}
where $\xi^{0} N^{-1}$ appears $n$ times. This leads to the
following expansion in $s$:
\begin{equation}
\xi_s^{0}(\theta) = \sum_{n=0}^{\infty} {1 \over
n!}\xi_s^{0\,[n]} s^{n}.  \label{xis}
\end{equation}
The one-parameter family of lapses $ N_s(\theta)$ follows almost
immediately from (\ref{xis}) and (\ref{dNdss}).
\begin{equation}
N_s(\theta) = N(\theta) + \sum_{n=0}^{\infty} {1 \over (n+1)!} {d \over
d\theta} \xi_s^{0\,[n]}(\theta) s^{n+1}. \label{ns}
\end{equation}

Let us also compute the one-parameter family of transformed
particle positions. Since the $x^\mu$ are scalars these families
will obey
\begin{equation}
{\partial x_s^\mu \over \partial s} = {\partial x_s^\mu \over
\partial \theta} \xi_s^0 N_s^{-1}. \label{parz}
\end{equation}
We find that
\begin{eqnarray}
{\partial^n x_s^\mu \over \partial s^n}|_{s=0}  &=& \xi^0 N^{-1}
{d \over \partial \theta }\left( \xi^0 N^{-1} {d \over
\partial \theta }\left( \xi^0 N^{-1}{d \over \partial \theta } \left(\dots {d \over \partial \theta }
\left(\xi^0 N^{-1}{d x^\mu \over \partial \theta }\right)
\ldots\right)
\right)\right) \nonumber \\
&=& \xi^0 N^{-1} {d \over \partial \theta }\left( \xi^0 N^{-1} {d
\over
\partial \theta }\left( \xi^0 N^{-1}{d \over \partial \theta } \left(\dots {d \over \partial \theta }
\left(\xi^0 p^\mu\right) \ldots\right) \right)\right),
 \label{zn}
\end{eqnarray}
where in the last line we used the equation of motion ${\dot
x^\mu} = N p^\mu$. The expression simplifies further using the
equation of motion ${\dot p^\mu} = 0$  yielding the formal
solution
\begin{equation}
x_s^\mu(\theta) = x^\mu(\theta)+p^\mu \sum_{n=0}^{\infty} {1 \over
n!} \xi_s^{0 [n]}(\theta) s^n. \label{xss}
\end{equation}

One special case for $\xi_s^{0}(\theta)$, $N_s(\theta)$ and
$x_s^\mu(\theta)$ is especially worthy of note. Notice that when
$\xi^{0}(\theta) = N(\theta)$ we have $\xi_s^{0}(\theta) =
\xi^{0}(\theta +s)$, $N_s(\theta) = N(\theta +s)$, and
$x_s^\mu(\theta) = x^\mu(\theta +s)$. The effect of this
one-parameter family of diffeomorphism-induced transformations on
the lapse and particle position is to advance them in $\theta$ by
the parameter value $s$! This is an illustration of a general
property of the one-parameter time-like diffeomorphism induced
transformations on the dynamical variables, evident also for the
explicit variations of lapse and shift exhibited in (\ref{d});
although the gauge group elements do not in general generate
rigid translations in time, when the group acts on solutions for
which the lapse $N$ and shift $N^{a}$ are equal to the
descriptors $\xi^{0}$ and $\xi^{a}$, respectively, the result is
to effect time evolution of the solutions. This is to be noted
for the infinitesmal transformations in (\ref{d}) since
substitution of these choices for the descriptors yields $\delta
N = \dot N ds$ and $\delta N^{a} = \dot N^{a} ds$.

We are finally able to write down the finite
diffeomorphism-induced generator of variations of $N$, $x^\mu$,
and $p^\mu$ in the free relativistic particle case. Letting
$G(\theta,s):= \int d\theta' \pi (\theta') \dot \xi_s^{0}(\theta')
+ \xi_s^{0}(\theta) \frac{1}{2} (\hat p_\mu \hat p^\mu +m^2)$,
the result is the parameter-ordered exponential
\begin{eqnarray}
    &&{\cal S}\left( exp^{\int_{0}^{s} ds'
    \{-,G(\theta,s') \}}\right) \nonumber \\
    && 1 + \int_{0}^{s} ds_{1}
     \{-, G(\theta,s_{1}) \}
    + \int_{0}^{s} ds_{2} \int_{0}^{s_{2}} ds_{1}
     \{ \{-,G(\theta,s_{1})\}, G(\theta,s_{2}) \} + \ldots \label{fg}
\end{eqnarray}

\section{The Quantum Physical Projector}

I shall elaborate here on a recent proposal
\cite{pons/salisbury/shepley/00} which is inspired by Carlo
Rovelli's introduction of an operator he calls a ``physical
projector'' \cite{rovelli/98}. The fundamental idea is to average
over the symmetry group. But before we can do this we must fix our
Hilbert space.

I propose to enlarge this Hilbert space to include
time-parametrized gauge functions in addition to the non-gauge
variables. If we take the non-gauge variables to be the spatial
metric, we obtain a generalization of the Wheeler-DeWitt approach
to quantum gravity. In this essay, however, I will generalize the
loop approach which employs the Ashtekar connection.

In part to explain the procedure, and also in part to check
whether it yields plausible results in a well understood simple
theory, we shall first address the group averaging question for
the free relativistic particle.

We consider a mixed momentum/Schr\"{o}dinger representation where
$\hat p^\mu$ and $\hat N(\theta)$ are multiplicative operators. We
interpret the argument $\theta$ of $\hat N(\theta)$ as a
parameter, so our quantum relativistic particle model has been
converted into a field theory. The momentum  conjugate to
$N(\theta)$ is thus also a field, as in our classical description
above.

Our task is to calculate the parameter-ordered
functional integral of the quantum version of the finite classical
generator given in (\ref{fg}). The corresponding physical projector is
\begin{equation}
    \hat{\cal P} :=  {\cal S}\left( [D \xi^{0}]
    exp\left(-i\int^{s}_{0} ds'
    \hat G( \theta,s')\right)\right)\ .
    \label{eq:path}
\end{equation}
Unfortunately, this is a highly non-trivial functional of the
descriptor field $\xi^{0}$, and a reasonable approximation scheme has
not yet been found for performing this functional integral.

\section{Loop Quantum Gravity}

The conventional approach to loop quantum gravity takes as the
kinematical arena a Hilbert space constructed from traces of
holonomies of closed spatial loops. (See \cite{gaul/rovelli/00}
for a recent review.)  These traces are invariant under SU(2)
gauge rotations of the holonomies. But as a consequence of
identities satisfied by the traces the states constructed with
them are linearly dependent. Rovelli and Smolin realized that the
isolation of a linearly independent set corresponded to the
notion of spin network that had been invented by Roger Penrose
\cite{rovelli/smolin/95b,penrose/71}.

I have not yet worked out all of the implications of the following
proposal, but it does seem to me to exhibit several attractive
features, and I would anticipate that some variation of it will
survive in a fully articulated quantum theory.

I want to retain the gauge variables as operators; the lapse
$N(\vec x,t)$, shift components $N^{a}(\vec x,t)$, and the time
components of the Ashtekar connection $A^{i}_{0}(\vec x,t)$ each
constitute a $4 \times \infty$ set of freely specifiable
variables. We can use them to construct spacetime loops with
associated holonomies. In particular I propose to attach timelike
legs to finite open paths in space. We might try, for example,
holonomies (parallel transport matrices) along timelike paths of
the form
\begin{equation}
        {\cal T}exp\left(\int_{t_{1}}^{t_{2}}dt' A_{\mu} n^{\mu}\right),
\end{equation}
where it is understood that $A_{a}$ are to be taken as
independent of $t$, and $A_{\mu} :=  A_{\mu}^{i} \tau_{i}$, the
$\tau_{i}$ being the Pauli matrices. Then we construct closed
spacetime loops, with associated holonomies, by first
transporting along a spatial path with fixed inital and final
points, say from $\vec x_{1}$ to $\vec x_{2}$, then forward in
time from $t_{1}$ to $t_{2}$, back along (a generally distinct)
spatial path from $\vec x_{2}$ to $\vec x_{1}$, and then finally
backward in time from $t_{2}$ to $t_{1}$. The trace of this
holonomy is invariant under internal $SU(2)$ rotations. Products
of the traces associated with loops will satisfy the same spinor
and retracing identities referred to above. We might reasonably
expect, therefore, that a four-dimensional spin network will
constitute a linearly independent set of kinematical states. This
hypothesis is now being explored, as are the following associated
problems: What is the relation of these states, after integrating
over the spatial diffeomorphism group, to the knot states in the
three-dimensional spin network formalism (see \cite{gaul/rovelli/00} for a review), and what is the
appropriate measure in this space?

The next task will be to attempt to give some sense to the formal
physical projector
\begin{equation}
\hat{\cal P} :=  {\cal S}\left( [D \xi]
    exp\left(-i\int^{s}_{0} ds'
    \hat G(t,s')\right)\right)\ .
    \label{Apath}
\end{equation}
This is a daunting challenge. On the one hand we do not yet have a
general expression for the one-parameter descripter families
$\xi^{\mu}(t,s)$. Even worse, the products of operators
appearing in $\hat G(t,s')$ are not well defined.
However, regularization techniques are available, and they have
been employed successfully in similar expansions, resulting in a
structure which has been called a ``spin foam''
\cite{thiemann/98,reisenberger/rovelli/97}. (Incidently, this
regularization technique, and the construction of the measure, are
achieved with a real Ashtekar connection. The symmetry generators
in this formulation also have the form (\ref{GGGGG})
\cite{pons/salisbury/00}.) The expansion and regularization of
(\ref{Apath}), and its relation to spin foams is the focus of
current research. Since spacetime area and volume operators will
very likely be well-defined in this formalism, we can reasonably
anticipate that when acting on four-dimensional networks we will
encounter eigenstates of these operators with discrete
eigenvalues.

\section{Conclusions}

In this essay I have reviewed our current understanding of
classical gauge symmetries in Hamiltonian formulations of
generally covariant theories which are derivable from a
Lagrangian. These symmetries form an infinite dimensional
transformation group, and I displayed explicitly the general form
of the generator of finite one-parameter subgroups which are
connected to the identity. Gauge functions are retained as
dynamical variables, and although they undergo non-trivial
variations under arbitrary symmetry transformations, their time
evolution is completely arbitrary. I have argued that there is
ample physical motivation for retaining these gauge variables in
a quantum theory of gravity. But recognizing their arbitrary
evolution, it is both reasonable and consistent with their
symmetry variations to conceive of their quantum operator
analogues as independent operators at  distinct times.

True symmetry invariants can then be obtained in this formalism by
integrating the finite quantum symmetry generator over the gauge
group. I call the resulting operator the physical projector.

I have proposed a tentative implementation of this approach in a
new loop approach to quantum gravity. Using the arbitrary gauge
functions in the Ashtekar approach which are the lapse, shift,
and the temporal component of the Ashtekar connection, we
construct traces of holonomies around spacetime loops. I
speculate that the resulting linearly independent states are
four-dimensional spin networks. The physical projector may be
expressed as an infinite expansion. Once the terms in this
expansion are suitably regularized, it may be possible either
truncate this expansion, or achieve partial infinite sums. We will then be in
position to calculate spacetime areas and volumes.

% xxxxx remove the following line before submitting
% \end{multicols}

\end{document}